\documentclass[twocolumn,tighten]{aastex63}

\usepackage{apjfonts}

\usepackage{amsmath}
\usepackage[T1]{fontenc}

\usepackage{CJK}
\newcommand{\Msun}{\,M_{\odot}}

\newcommand{\be}{\begin{equation}}
\newcommand{\ee}{\end{equation}}
\newcommand{\bea}{\begin{eqnarray}}
\newcommand{\eea}{\end{eqnarray}}
\def\kms{\ {\rm km\, s}^{-1}}

\def\msun{M_\odot}

\shortauthors{CONROY ET AL.}
\shorttitle{Ark 227}

\begin{document}

%---------------------------------------------------------%

\title{Detection of Accretion Shelves Out to the Virial Radius of a
  Low-Mass Galaxy with {\it JWST}}

\author{Charlie Conroy}
\affiliation{Center for Astrophysics $\mid$ Harvard \& Smithsonian, 60 Garden St, Cambridge, MA 02138, USA}

\author{Benjamin D. Johnson}
\affiliation{Center for Astrophysics $\mid$ Harvard \& Smithsonian, 60 Garden St, Cambridge, MA 02138, USA}

\author{Pieter van Dokkum}
\affiliation{Department of Astronomy, Yale University, New Haven,
 CT, 06511, USA}

\author{Alis Deason}
\affiliation{Institute for Computational Cosmology, Department of Physics, Durham University, Durham DH1 3LE, UK}

\author{Sandro Tacchella}
\affiliation{Kavli Institute for Cosmology, University of Cambridge, Madingley Road, Cambridge, CB3 0HA, UK}
\affiliation{Cavendish Laboratory, University of Cambridge, 19 JJ Thomson Avenue, Cambridge, CB3 0HE, UK}

\author{Sirio Belli}
\affiliation{Dipartimento di Fisica e Astronomia, Università di
  Bologna, Bologna, Italy}

\author{William P. Bowman}
\affiliation{Department of Astronomy, Yale University, New Haven,
 CT, 06511, USA}

\author{Rohan P. Naidu}
\altaffiliation{NASA Hubble Fellow}
\affiliation{MIT Kavli Institute for Astrophysics and Space Research, 77 Massachusetts Ave., Cambridge, MA 02139, USA}

\author{Minjung Park}
\affiliation{Center for Astrophysics $\mid$ Harvard \& Smithsonian, 60 Garden St, Cambridge, MA 02138, USA}

\author{Roberto Abraham}
\affiliation{Department of Astronomy \& Astrophysics, University of Toronto, 50 St. George Street, Toronto, ON M5S 3H4, Canada}

\author{Razieh Emami}
\affiliation{Center for Astrophysics $\mid$ Harvard \& Smithsonian, 60 Garden St, Cambridge, MA 02138, USA}

\begin{abstract}

  We report the serendipitous discovery of an extended stellar halo
  surrounding the low-mass galaxy Ark 227 ($M_\ast=5\times10^9\msun$;
  $d=35$ Mpc) in deep {\it JWST} NIRCam imaging from the Blue Jay
  Survey.  The $F200W-F444W$ color provides robust star-galaxy
  separation, enabling the identification of stars at very low
  density.  By combining resolved stars at large galactocentric
  distances with diffuse emission from NIRCam and Dragonfly imaging at
  smaller distances, we trace the surface brightness and color
  profiles of this galaxy over the entire extent of its predicted dark
  matter halo, from $0.1-100$ kpc.  Controlled N-body simulations have
  predicted that minor mergers create ``accretion shelves'' in the
  surface brightness profile at large radius.  We observe such a
  feature in Ark 227 at $10-20$ kpc, which, according to models, could
  be caused by a merger with total mass ratio 1:10.  The metallicity
  declines over this radial range, further supporting the minor merger
  scenario.  There is tentative evidence of a second shelf at
  $\mu_V\approx 35$ mag arcsec$^{-2}$ extending from $50-100$ kpc,
  along with a corresponding drop in metallicity.  The stellar mass in
  this outermost envelope is $\approx10^7\msun$.  These results
  suggest that Ark 227 experienced multiple mergers with a spectrum of
  lower-mass galaxies -- a scenario that is broadly consistent with
  the hierarchical growth of structure in a cold dark matter-dominated
  universe.  Finally, we identify an ultra-faint dwarf associated with
  Ark 227 with $M_\ast\approx10^5\msun$ and $\mu_{V,e}=28.1$ mag
  arcsec$^{-2}$, demonstrating that {\it JWST} is capable of detecting
  very low-mass dwarfs to distances of at least $\sim30$ Mpc.
  
\end{abstract}


%---------------------------------------------------------%

\section{Introduction}
\label{s:intro}

The cold dark matter (CDM) cosmological model predicts that structure
forms ``bottom-up'', in which larger, more massive dark matter halos
grow from the assimilation of many smaller halos.  This process is
most clearly observed in the spectacular faint tidal features and
complex stellar halos observed in the Milky Way and other galaxies of
comparable or greater masses \citep[e.g.,][]{Majewski03, Mihos05,
  Belokurov06, Martinez-Delgado10, McConnachie09, Duc15, Merritt16,
  Naidu20a}.

CDM predicts that this assimilation process is approximately
scale-free, such that dwarf mass halos should form from the accretion
of still lower mass objects.  In fact, observations of dwarf halos may
even provide a probe of the nature of dark matter on these scales
\citep{Deason22}.  However, the relation between galaxy mass and dark
matter halo mass is very steep and uncertain at low masses.  It is
therefore unclear if the predicted bottom-up, accretion-driven,
formation process is observable in the form of tidal debris and
stellar halos at the scale of dwarf galaxies.  Observations of a tidal
stream around the dwarf galaxy NGC 4449 provide the sole unambiguous
example of hierarchical assembly at the dwarf scale
\citep[][]{Strader12, Martinez-Delgado12}.

Hydrodynamic simulations predict that in-situ processes associated
with bursty stellar feedback can drive stars born in dwarf galaxies to
halo-like orbits \citep[e.g.,][]{El-Badry16, Kado-Fong22}.  The bursty
feedback is more common at early times, when the metallicity was
lower.  The stars on halo-like orbits therefore tend to be lower
metallicity than the inner regions.  The existence of metal-poor stars
in the outskirts of dwarf galaxies is not in itself evidence of the
hierarchical assembly process acting at the dwarf scale.  This
alternative channel for stellar halo formation has frustrated efforts
to interpret stellar populations in the outskirts of dwarf galaxies
\cite[e.g.,][]{Chiti21}.

The spatial distribution of the most distant halo stars surrounding
dwarf galaxies may break these degeneracies.  Hierarchical merger
models predict accretion shelves in the surface brightness profiles of
dwarfs that are directly related to the mass ratio of the merger
\citep{Amorisco17, Deason22}.  Furthermore, in-situ models seem unable
to populate stars to a large fraction of the virial radius, unlike
accretion scenarios \citep[e.g.,][]{Kado-Fong22}.  Surface brightness
measurements of a dwarf galaxy to its virial radius should therefore
provide strong constraints on the physical origin of its stellar halo.

In this paper we present serendipitous observations of the stellar
halo of the galaxy Ark 227 observed by {\it JWST} as part of the Blue
Jay Survey in the COSMOS field (Belli et al., in prep).  The
sensitivity of {\it JWST} imaging allows us to trace the stellar halo
of this galaxy to its dark matter halo virial radius.  The observed
surface brightness and metallicity profiles suggest that this galaxy's
stellar halo was built from the assimilation of smaller-mass galaxies,
providing a dramatic example of hierarchical assembly at the dwarf
scale.

Magnitudes reported in this paper adopt the AB zero point system
\citep{Oke83}.

%--------------------------------------------------------%
%--------------------------------------------------------%
%--------------------------------------------------------%

\section{Data \& Methods}
\label{s:data}

\subsection{{\it JWST} imaging and ``discovery'' of Ark 227}

The Blue Jay Survey is a Cycle 1 {\it JWST} program (GO 1810; PI
Belli).  The primary scientific objective of the program is to obtain
deep spectra of a mass-selected sample of galaxies at cosmic noon
($1.7<z<3.5$).  The NIRSpec micro-shutter array was used to obtain
$R\approx1000$ spectra of 150 galaxies with three medium resolution
gratings over two separate pointings.  Parallel observations were
obtained with NIRCam in a variety of filters of varying depths.  The
filters and associated exposure times are: $F090W$ (92 min), $F115W$
(184 min), $F150W$ (368 min), $F200W$ (368 min), $F277W$ (276 min),
$F356W$ (368 min), and $F444W$ (368 min).  Details of the program will
be described in Belli et al. (in prep).

The orientation constraints of the NIRSpec observations resulted in
one of the NIRCam modules being placed south-west of the available
{\it HST}/CANDELS data in the COSMOS field.  To our surprise, a nearby
dwarf galaxy also happens to reside just south-west of the
CANDELS-COSMOS field.  We initially viewed this foreground dwarf with
some dismay, as the imaging in this module is badly ``contaminated''
with not only the unresolved light from Ark 227 but also the resolved
starlight associated with its stellar halo.  While we were visually
inspecting the mosaic far from Ark 227 we noticed a large number of
point sources with a common color (which happened to be green in the
adopted color map).  Tracing these green dots across the mosaic, we
realized that they appeared to be associated with Ark 227.  As we will
argue below, these green dots are the stellar halo of Ark 227 and are
present throughout the entire NIRCam mosaic.

Ark 227 \citep[PGC 28923;][]{Arakelian75} is a dwarf galaxy with red
colors and elliptical morphology.  Its redshift is $1793\kms$.  There
is no reliable distance measurement in the literature.
\citet{Leroy19} adopt a Hubble flow-based distance of 26 Mpc to infer
a stellar mass of $2.7\times10^9\msun$.  We present a (much more
accurate) TRGB-based distance of 35 Mpc below, with a corresponding
larger stellar mass.  Adopting the stellar mass-halo mass relation
from \citet{Behroozi13} implies a halo mass of $2\times10^{11}\msun$
and hence a virial radius of $\approx100$ kpc.  Galactic extinction
toward Ark 227 is small; we adopt $E(g-r) = 0.016$ from
\citet{Schlafly11}.  Its environment has not been studied in detail,
but it is not known to be associated with any bright galaxy
\citep{Polzin21}.

\begin{figure*}[!t]
\center
\includegraphics[width=1.0\textwidth]{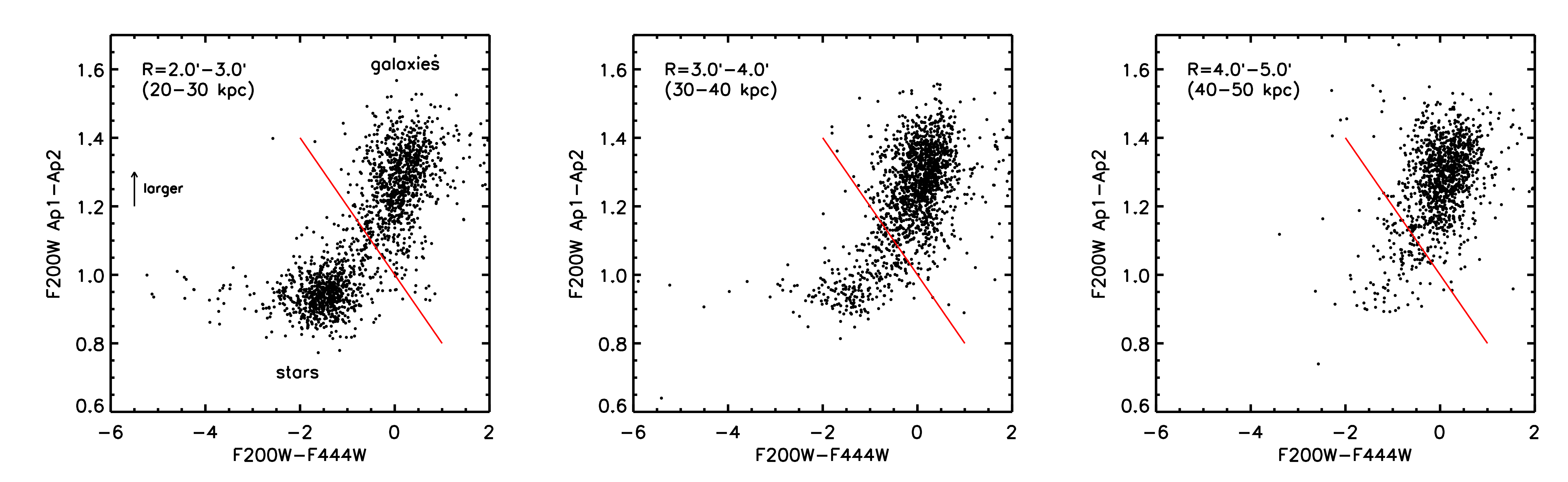}
\vspace{0.1cm}
\caption{``Pseudo-size'' vs. color diagrams at increasing distance
  from the center of Ark 227.  The $F200W-F444W$ color provides strong
  separation between stars and galaxies and is measured in fixed
  apertures of two pixel radii.  The difference in $F200W$ magnitudes
  measured in a one and two pixel aperture ($0.03\arcsec$ and
  $0.06\arcsec$; $Ap1-Ap2$) is shown on the y-axis and is a proxy for
  the size of the source.  The red line is our initial selection of
  star-like objects.  Sources below this line are passed through the
  light profile fitting algorithm \texttt{forcepho}.}
\label{fig:col-size}
\end{figure*}

\subsection{Data reduction and photometry}

The imaging used a 4-point dither pattern dictated by the
spectroscopic program. The individual exposures were processed using
the `jwst' data reduction pipeline version 1.9.4 with CRDS context map
`jwst\_1039.pmap'.  The exposures were then astrometrically aligned
using a reference catalog derived from {\it HST} imaging registered to
{\it Gaia} \citep{GaiaDR2, Mowla19} before final mosaicing.  Sources
were detected on a stack of the $F150W$ and $F200W$ mosaics via the
Source Extractor program \citep{Bertin96}, and circular aperture
photometry was measured in all bands for the sources at the detection
coordinates in several apertures, including one and two pixel radii. A
subset of sources (described below) were passed through the
\texttt{forcepho} program (Johnson et al., in prep), which fits
PSF-convolved Sersic profiles to the multiband exposure level images
of each source, enabling measurements of source sizes (half-light
radii), colors, and total integrated fluxes.

\subsection{Selection of stars}

We experimented with a variety of diagnostic diagrams in order to
separate stars from galaxies.  The most useful diagnostic is a
combination of short and long wavelength photometry (in particular,
$F200W-F444W$) and a proxy for the object size \citep[see also][who
advocate a short and long wavelength {\it JWST} filter for efficient
star-galaxy separation]{Warfield23}.  For the latter, we adopted the
difference between one and two pixel aperture ($0.03\arcsec$ and
$0.06\arcsec$) photometry in the $F200W$ band ($Ap1-Ap2$).  This band
is our deepest and so has the best SNR.  For point sources, the
difference in aperture photometry is a measure of the point spread
function and hence should be approximately a constant.  Stars have
blue colors in restframe $F200W-F444W$ because both filters are
redward of the $1.5\mu m$ peak of cool stars.  Galaxies are
intrinsically redder in this color because they are a composite
stellar population that includes very cool stars.  However, a larger
effect is redshifting: at $z>0.3$ the $F200W$ filter is sampling flux
blueward of the $1.5\mu m$ SED peak, which results in substantially
redder $F200W-F444W$ colors.

The resulting diagnostic diagram is shown in Figure \ref{fig:col-size}
at three annuli of increasing distance from Ark 227.  This diagram
clearly shows two distinct populations: compact sources at
$F200W-F444W\approx-1.75$ and extended sources at
$F200W-F444W\gtrsim-1$.  The relative proportion of sources in the
blue and red loci change markedly from the inner to outer regions of
Ark 227, strongly suggesting that the blue sources are stars
associated with Ark 227.

\begin{figure}[!t]
\center
\includegraphics[width=0.45\textwidth]{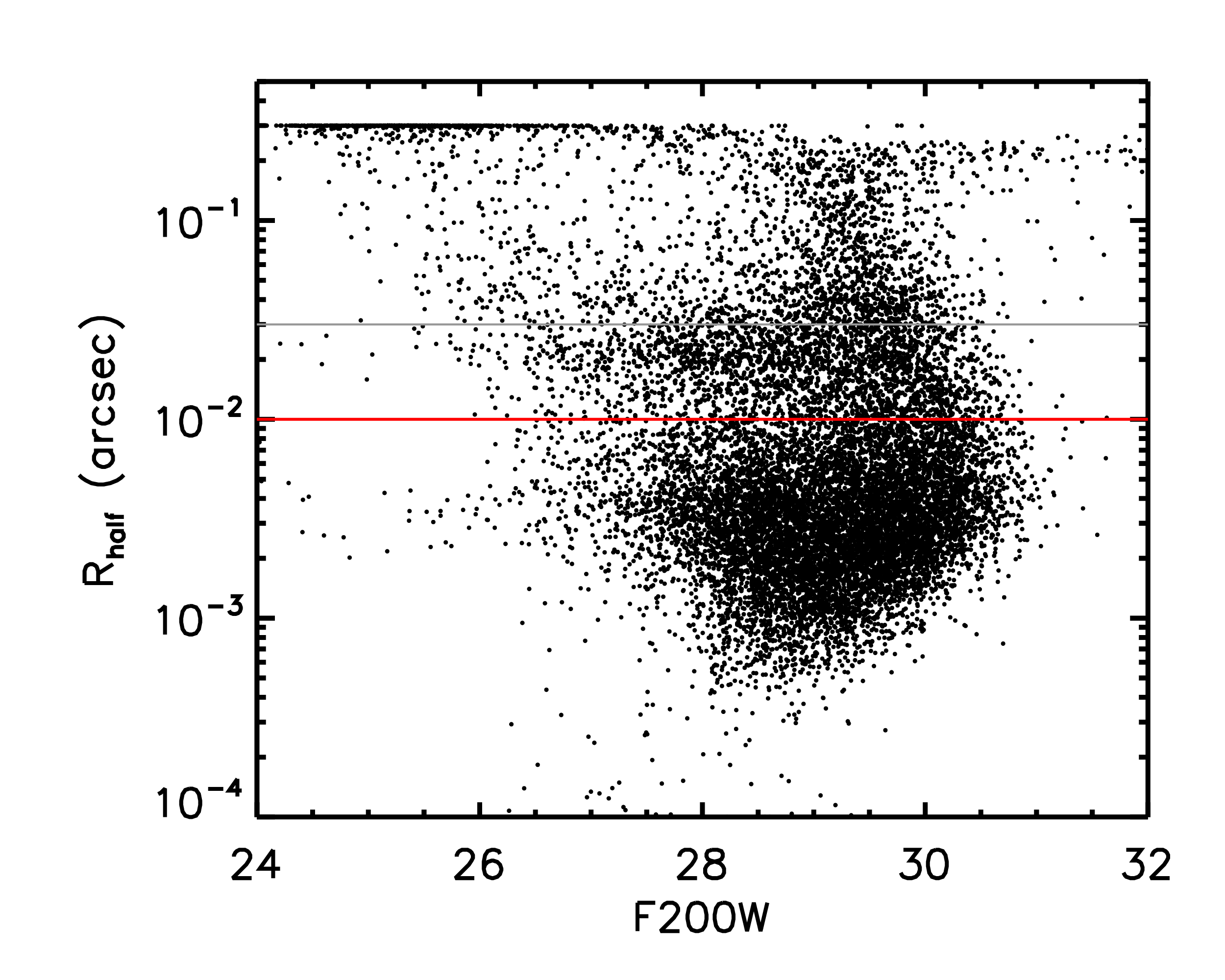}
\caption{Size vs. $F200W$ magnitude for all sources selected as
  ``star-like'' in Figure \ref{fig:col-size}.  Sizes and magnitudes
  are determined from profile fitting.  The grey line is the size of a
  NIRCam pixel in the SW module. Our sample of star-like point
  sources is comprised of all objects below the red line. }
\label{fig:size-mag}
\end{figure}

\begin{figure*}[!t]
\center
\includegraphics[width=1.0\textwidth]{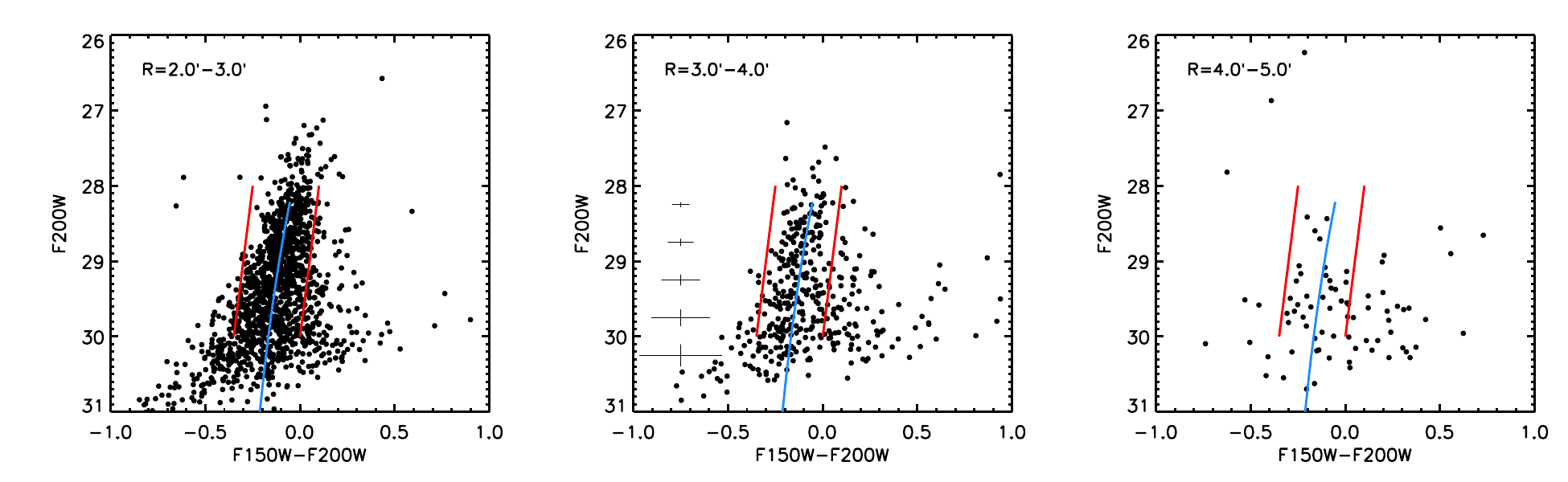}
\caption{Color-magnitude diagrams of star-like point sources at
  increasing separation from the center of Ark 227.  Red lines
  demarcate our final selection of stars.  A 10 Gyr [Fe/H]$=-1$
  \texttt{MIST} isochrone is shown at a distance of 35 Mpc (blue
  lines).  Typical uncertainties are shown in the middle panel as a
  function of magnitude. }
\label{fig:cmd}
\end{figure*}

\begin{figure}[!t]
\center
\includegraphics[width=0.45\textwidth]{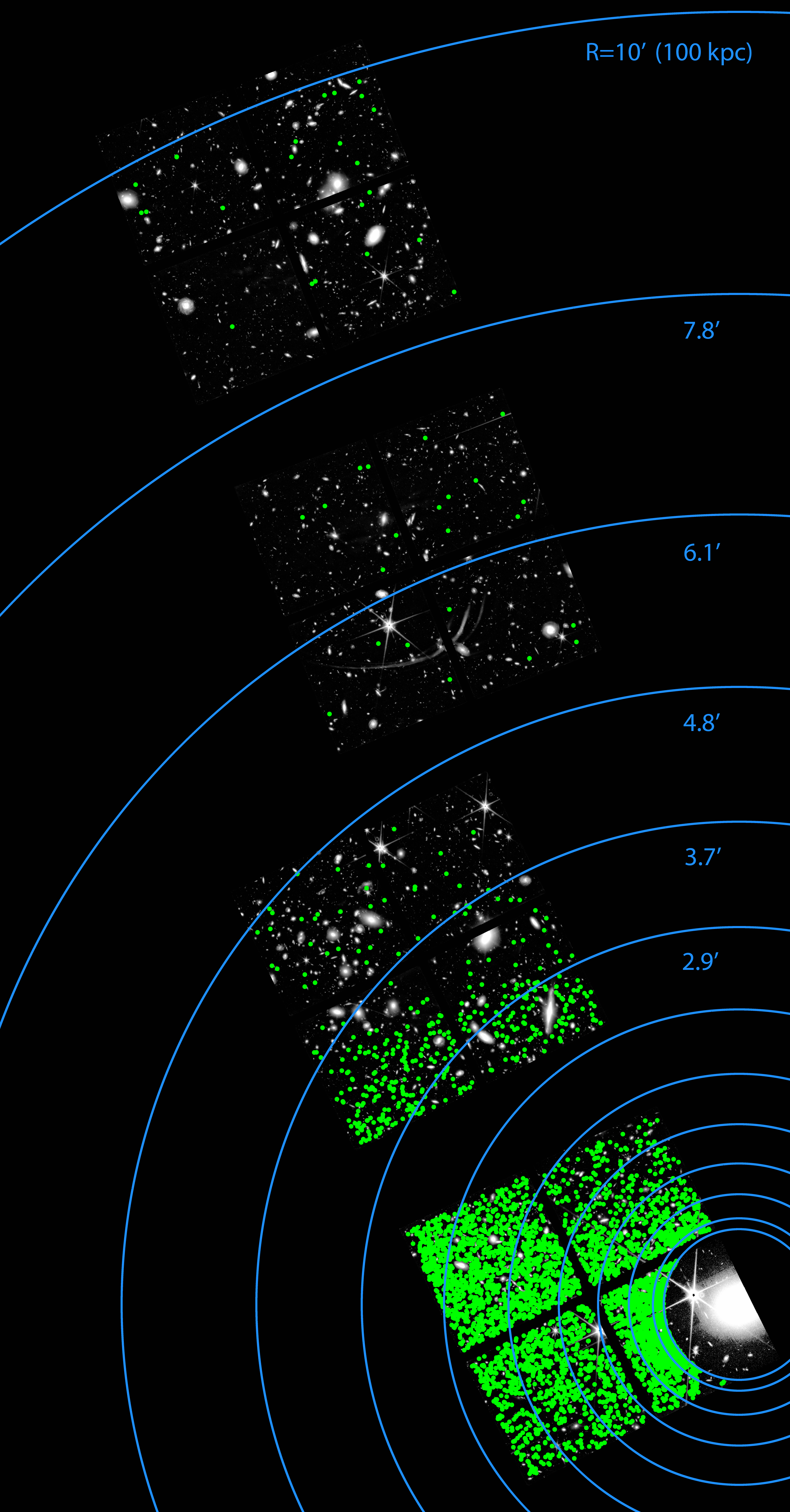}
\vspace{0.1cm}
\caption{Mosaic of the Blue Jay data in the $F200W$ filter.  Ark 227
  is clearly visible in the lower right corner.  Annuli from which
  surface brightness measurements are made via resolved stars are
  shown in blue.  Our final sample of star candidates associated with
  Ark 227 is shown as green points.}
\label{fig:layout}
\end{figure}

Our first selection of stars consists of sources falling below the red
line in Figure \ref{fig:col-size}; specifically:
$Ap1-Ap2<-0.2(F200W-F444W)+1.0$.  This is a generous selection,
including a number of sources that are likely not stars.  All objects
satisfying this selection are passed through the profile fitting
program \texttt{forcepho}.

Figure \ref{fig:size-mag} shows the half-light radius vs. total
$F200W$ magnitude for all objects satisfying the star-like selection
in Figure \ref{fig:col-size}.  The {\it JWST} NIRCam pixel size is
$0.03\arcsec$ in the SW module; this scale is included as a grey line
in Figure \ref{fig:size-mag}.  Most objects are very compact and are
effectively unresolved.  We adopt a selection of
$R_{\rm half}<0.01\arcsec$ to isolate stars from resolved objects.

\begin{figure}[!t]
\center
\includegraphics[width=0.45\textwidth]{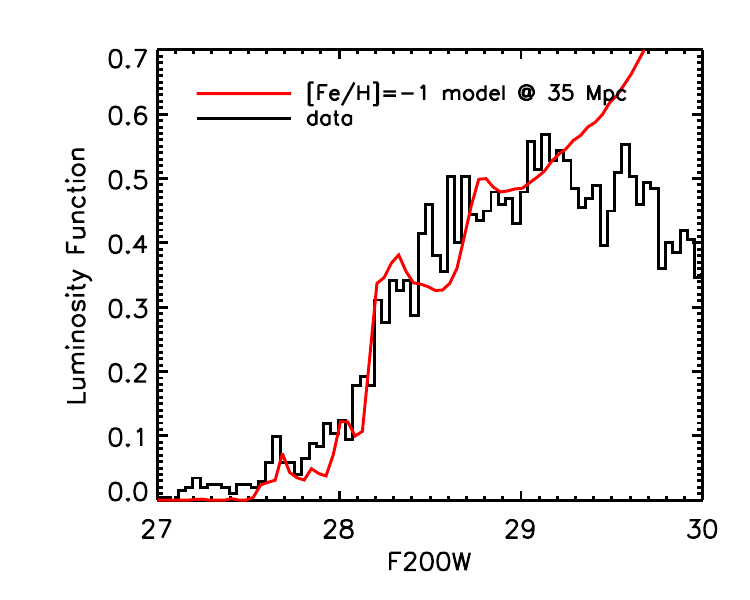}
\caption{Luminosity function (LF) of stars associated with Ark 227.
  The red line shows a model LF assuming a distance of 35 Mpc from the
  \texttt{MIST} isochrones.}
\label{fig:trgb}
\end{figure}

Figure \ref{fig:cmd} shows a color-magnitude diagram (CMD) in $F150W$
and $F200W$ filters for the star-like point sources passing the
selection criteria defined in Figures \ref{fig:col-size} and
\ref{fig:size-mag}.  The CMDs are shown in three annuli at increasing
distance from Ark 227.  A 10 Gyr model red giant branch at
[Fe/H]$=-1.0$ is overplotted as a blue line at an assumed distance of
35 Mpc.  Average uncertainties are shown as a function of magnitude in
the middle panel.  The CMD of these star-like point sources is clearly
consistent with the evolved giants of an old metal-poor stellar
population.

Our final selection of stars associated with Ark 227 consists of the
objects bounded by the nearly vertical red lines in the CMD,
restricted to $28<F200W<30$.  The color selection ensures that the
stars have RGB-like colors.  The bright limit rejects very bright AGB
stars but also limits potential interlopers, such as globular clusters
and foreground (MW) dwarf stars.  The faint limit is set by the
photometric depth.

In summary, our selection for stars associated with Ark 227 is based
on the selections shown in Figures \ref{fig:col-size},
\ref{fig:size-mag}, and \ref{fig:cmd} and is designed to select point
sources with RGB colors.  The resulting distribution of stars is shown
in Figure \ref{fig:layout}, along with the F200W mosaic.  Also shown
in this figure are the circular annuli within which the surface
brightness will be measured.

The CMD shows a sharp tip of the RGB (TRGB) location at
$F200W\approx28.2$.  To explore this further we show the stellar
luminosity function in Figure \ref{fig:trgb}.  We include a comparison
to a model luminosity function of a 10 Gyr, [Fe/H]$=-1.0$ stellar
population from the \texttt{MIST} isochrones.  We have placed the
model at 35 Mpc and added magnitude-dependent uncertainties comparable
to the data.  The distance was fit by-eye; more sophisticated fitting
is not warranted given that the uncertainty on the distance is
dominated by the dependence of the RGB-tip on metallicity: for [Fe/H]
$=-0.75, -1.0, -1.5$ the RGB-tip is $F200W=-4.73, -4.55, -4.23$.  An
uncertainty on the distance modulus of 0.25 mag corresponds to a
distance uncertainty of 12\%.  We therefore adopt a distance to Ark
227 of $35\pm4$ Mpc.

\subsection{Surface brightness measurements}
\label{s:sb}

Our primary goal is to measure the surface brightness profile of Ark
227.  At $R<10\arcsec$ we directly measure the diffuse light of Ark
227 from {\it JWST} NIRCam imaging.  At $R>35\arcsec$ we measure
resolved star photometry of Ark 227.  In the latter case, we convert
the observed flux measured over the magnitude range $28< F200W <30$ to
an integrated flux.  We do this by employing a 10 Gyr [FeH] $=-1$
isochrone from v2.3 of the \texttt{MIST} models \citep{Choi16}.  The
fraction of light in the observed magnitude range is $26$\%; we use
this fraction to correct our observed flux to an estimate of the total
flux (this fraction varies from $23-26$\% over the range
$-1.5<$[Fe/H]$<-0.5$).  In order to convert our total $F200W$ fluxes
to the more commonly-reported $V-$band flux, we adopt an integrated
color of $V-F200W=0.74$, appropriate for a 10 Gyr, [FeH]$=-1$
isochrone.

As discussed in Section \ref{s:ufd}, we identified an ultra-faint
dwarf galaxy in the halo of Ark 227.  The stars associated with this
dwarf are removed before measuring the surface brightness profile of
Ark 227.

We assess the level of contamination in our star selection in several
ways.  First, we use an empirically-calibrated model of star counts in
the Milky Way \citep{Girardi05} to estimate contamination from
foreground stars.  We find a likely contamination from Milky Way stars
that pass our CMD selection to be at the level of $\mu_V\approx 38$
mag arcsec$^{-2}$.

Second, we employ data from the JADES Survey \citep{Eisenstein23},
which achieves photometric depths comparable to Blue Jay.  We select
three $\approx 10$ arcmin$^2$ regions and use aperture photometry to
select star-like candidates.  We apply the same CMD selection and
magnitude limits as used in the Blue Jay data.  Since
\texttt{forcepho} photometry is not available, we employ a stricter
selection in the pseudo-size-color diagram.  Specifically, we select
stars in a box defined by $-3.0<F200W-F444W<-0.5$ and
$0.85<(Ap1-Ap2)<1.0$.   If we analyze the sources identified this way
in the same manner as the Ark 227 sources, we arrive at surface
brightness limits of $\mu_V=37.4, 36.1, {\rm and}\ 35.9$ mag arcsec$^{-2}$.

Inspection of the images reveals that many of the star-like sources
identified in JADES are associated with nearby ($z\lesssim0.2$) bright
galaxies.  It is likely that these sources are unresolved massive
globular clusters.  The JADES fields with a greater number of
spectroscopically-confirmed galaxies at $z<0.2$ correspond to fields
with higher densities of star-like sources, further supporting this
conclusion.  We therefore regard extragalactic globular clusters as
the dominant source of contamination in our analysis.  Their visible
association with bright galaxies should enable their identification
and removal.  Though we leave a full treatment of this next step to
future work, we visually inspected the Blue Jay data and found that in
the outermost radial bin ($>7.8\arcmin$) as many as 50\% of the
sources could plausibly be associated with foreground galaxies.  We
took a conservative approach and removed 50\% of the sources in this
last radial bin.  In the next two radial bins
($4.8\arcmin-7.8\arcmin$) we saw little evidence for associations and
so made no corrections.  Owing to these complications, we regard
measurements of surface brightness at $\mu_V\gtrsim 35$ mag
arcsec$^{-2}$ as tentative.

\begin{figure*}[!t]
\center
\includegraphics[width=0.95\textwidth]{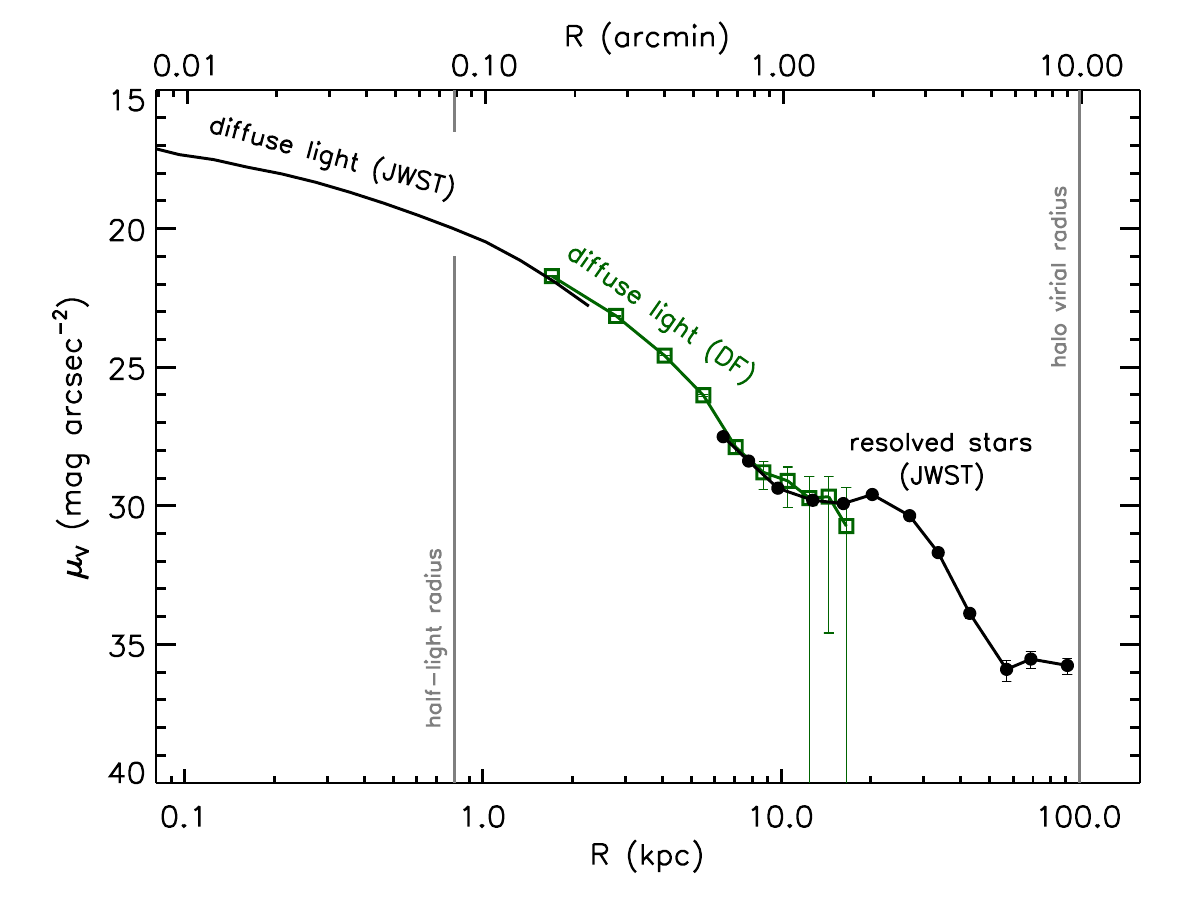}
\caption{Surface brightness profile of Ark 227 from $0.1-100$ kpc,
  measured via diffuse light from {\it JWST} NIRCam imaging, diffuse
  light from Dragonfly (DF) imaging, and resolved starlight from {\it
    JWST} NIRCam photometry.  The half-light and halo virial radii are
  indicated with vertical grey lines.  A TRGB-based distance of 35 Mpc
  was adopted to convert angles into projected distances.}
\label{fig:sb1}
\end{figure*}

\subsection{Dragonfly imaging}

At high surface densities it is relatively straightforward to measure
the diffuse emission from a galaxy, while at very low surface
densities, where stellar crowding is not a concern, it is conventional
to measure surface brightness from resolved star counts.  At the
boundary of these two regimes challenges abound.  At one end, crowding
makes resolved star measurements increasingly challenging.  At the
other end, the diffuse flux level is so faint that special techniques
and/or observatories are required to avoid systematic errors from
scattered light, flat fielding, etc.  Our current reduction of the
{\it JWST} NIRCam data does not deliver reliable diffuse flux
measurements beyond $R\approx10\arcsec$.  A 12$^{\rm th}$ magnitude
star resides $22\arcsec$ from the center of Ark 227, further
complicating efforts at measuring low surface brightness features via
diffuse emission.

To bridge the resolved and diffuse emission regimes we turn to the
Dragonfly Telephoto Array, a special-purpose observatory designed for
low surface brightness imaging \citep{Abraham14}. Ark227 is close in
projection to the unrelated edge-on spiral galaxy NGC 3044 at a
distance of 23 Mpc, and it was serendipitously observed in a deep
Dragonfly observation of that galaxy. The Dragonfly data for NGC 3044
are described in \citet{Gilhuly22}; they reach $3\sigma$ surface
brightness limits of $29.8$ mag arcsec$^{-2}$ in $g$ and
$29.1$ mag arcsec$^{-2}$ in $r$.  They were re-reduced with the latest
Dragonfly pipeline (Bowman et al., in prep).

Stars and other compact objects were subtracted from the Dragonfly
data with the multi-resolution filtering technique
\citep{vanDokkum20d}, using Legacy Survey imaging as input to the
model \citep[see][]{Gilhuly22}.  The 12$^{\rm th}$ magnitude star is
saturated in the Legacy imaging but not in the Dragonfly data; it was
subtracted with a custom wide-angle PSF created from other bright
stars in the field. Residuals of bright subtracted objects were
masked, as described in \citet{vanDokkum20d}.  Surface brightness
profiles in $g$ and $r$ were measured from the filtered image using
aperture photometry, taking missing flux in masked regions properly
into account.  We find no evidence for asymmetries in the light
distribution, but we note that this is difficult to assess in the
vicinity of the bright star and other relatively bright stars to the
North of Ark227.  Uncertainties were determined from the empirical
variations in the background in empty areas of the images.

%--------------------------------------------------------%
%--------------------------------------------------------%
%--------------------------------------------------------%

\section{Results}
\label{s:res}

\subsection{Surface brightness profile}

Figure \ref{fig:sb1} shows the final surface brightness profile for
Ark 227 from $0.1-100$ kpc.  We combine diffuse light measurements
from {\it JWST} NIRCam and Dragonfly (DF) imaging with estimates from
resolved star counts at large radius.  The half-light radius and
approximate dark matter halo virial radius are indicated with grey
lines.  The agreement between the three distinct tracers of the
surface brightness profile is excellent and provides a good check that
none of the probes contains serious systematic errors.  The surface
brightness profile spans 18 magnitudes, for a total change in
intensity of $\approx 10^7$.

There are three distinct regimes in the surface brightness profile.
At $R\lesssim6$ kpc the profile is smoothly declining and is
reasonably well-described by a single Sersic model.  At
$10\lesssim R\lesssim 50$ kpc the surface brightness profile flattens
to a level of $\mu_V\approx 30$ mag arcsec$^{-2}$ and then drops
rapidly over $20\lesssim R\lesssim 50$ kpc. We refer to this
morphology as a shelf in the brightness profile.  Finally, at
$R\gtrsim 50$ kpc the surface brightness flattens again, with no sign
of truncation to the limit of our data (although measurements at
$\mu\gtrsim35$ mag arcsec$^{-2}$ are tentative for reasons discussed
in Section \ref{s:sb}).  Assuming a stellar mass-to-light ratio of
$M/L_V=2$, the mass within 10 kpc is $5.2\times10^9\Msun$, within
$10<R<50$ kpc is $2.0\times10^8\Msun$, and at $>50$ kpc is
$1.0\times10^7\msun$.  We return to potential origins of these
distinct regimes below.

\begin{figure*}[!t]
\center
\includegraphics[width=1.0\textwidth]{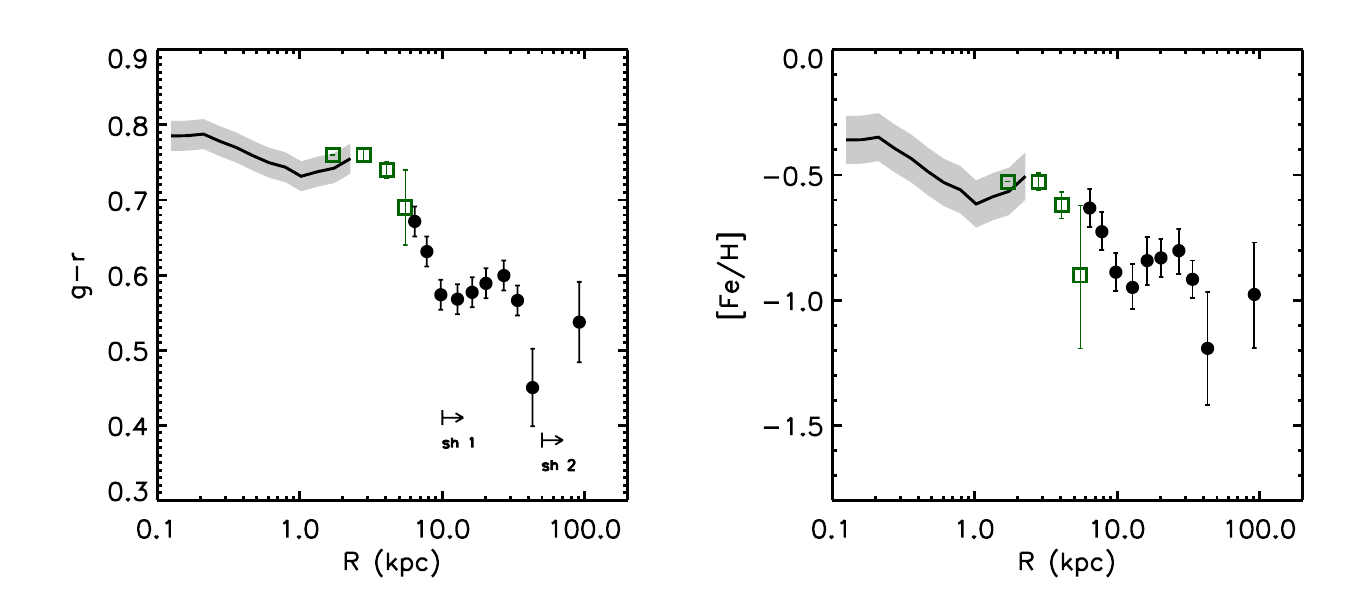}
\caption{{\it Left panel:} Color profile of Ark 227.  The diffuse
  emission measured from NIRCam imaging (solid line) is converted to
  $g-r$ from the observed $F115W-F200W$ color assuming a color
  conversion of 0.7 mag.  The diffuse emission measured from Dragonfly
  imaging (green squares) is obtained in $g$ and $r$ filters.  The
  resolved star data is converted from a mean RGB color to an
  integrated $g-r$ based on isochrones.  {\it Right panel:}
  Metallicity profile estimated from the measured diffuse and resolved
  colors.  The color and metallicity abruptly decrease at the
  locations where the surface brightness profile flattens, indicated
  by the arrows and `sh 1', `sh 2', in the left panel. }
\label{fig:colfeh}
\end{figure*}

\subsection{Color and metallicity profile}

The left panel of Figure \ref{fig:colfeh} shows the $g-r$ color
profile of Ark 227.  At the smallest radii the color is estimated
directly from the diffuse emission in {\it JWST} NIRCam imaging using
the $F115W$ and $F200W$ filters.  We then adopt a fixed color term of
$g-r=F115W-F200W+0.70$ based on integrated colors from 10 Gyr
isochrones.  This color term is nearly constant over a wide range in
metallicity, varying by $\pm0.05$ over $-2<$[Fe/H]<$+0.5$.  The
Dragonfly data were obtained in $g$ and $r$ filters and so $g-r$
colors can be readily measured from those data.  At the largest
distances where resolved star data are employed, we use a 10 Gyr
[Fe/H]$=-1$ isochrone to determine an offset between the RGB
$F115W-F200W$ color and the integrated $g-r$ color of 0.36.  This is
only approximate because the color variation seen in the data is
likely a reflection of underlying metallicity variation, and the mean
luminosity of stars changes slightly with distance.  

The right panel of Figure \ref{fig:colfeh} shows the estimated stellar
metallicity profile of Ark 227.  For the integrated light measurements
we use color-metallicity relations from isochrones to translate the
observed $F115W-F200W$ and $g-r$ colors into metallicities.  For the
Dragonfly $g-r$ data, we have applied a small offset of $-0.03$ in the
color before converting to metallicities in order to provide a
slightly better match to the metallicity at smaller scales.  This
offset could be due to zero point uncertainties in the photometry, or
small offsets in the color-metallicity relations in different bands.
For the resolved stars, we compute the mean luminosity in each annulus
and construct an RGB $F115W-F200W$ color vs. metallicity relation at
the associated mean luminosity.  We then use the observed color to
estimate a metallicity.

Precise metallicities will require either spectroscopy (which will be
very challenging at these depths) or additional photometric bands.
The key takeaway from Figure \ref{fig:colfeh} is that the shelves in
the surface brightness profile correspond to abrupt declines in the
color and therefore metallicity profiles.  This strongly suggests that
the distinct features in the light profile correspond to distinct
stellar populations.

\subsection{Comparison to hierarchical growth models}

In this section we compare our results to cosmologically-motivated
merger models.  We employ the framework presented in \citet{Deason22}
in which idealized dark matter-only mergers are run with the
\texttt{GADGET-2} code.  Stars are assigned to dark matter halos
according to a stellar mass-halo mass relation and a particle tagging
technique with an observationally-motivated size-mass relation.
Deason et al. focused on the merger histories of dwarfs with halo
masses $10^{10}\msun$.  Here we consider models more appropriate for
Ark 227 with $M_{\rm halo}=10^{11}\msun$ and a halo concentration of
$c=10$.

\begin{figure*}[!t]
\center
\includegraphics[width=1.0\textwidth]{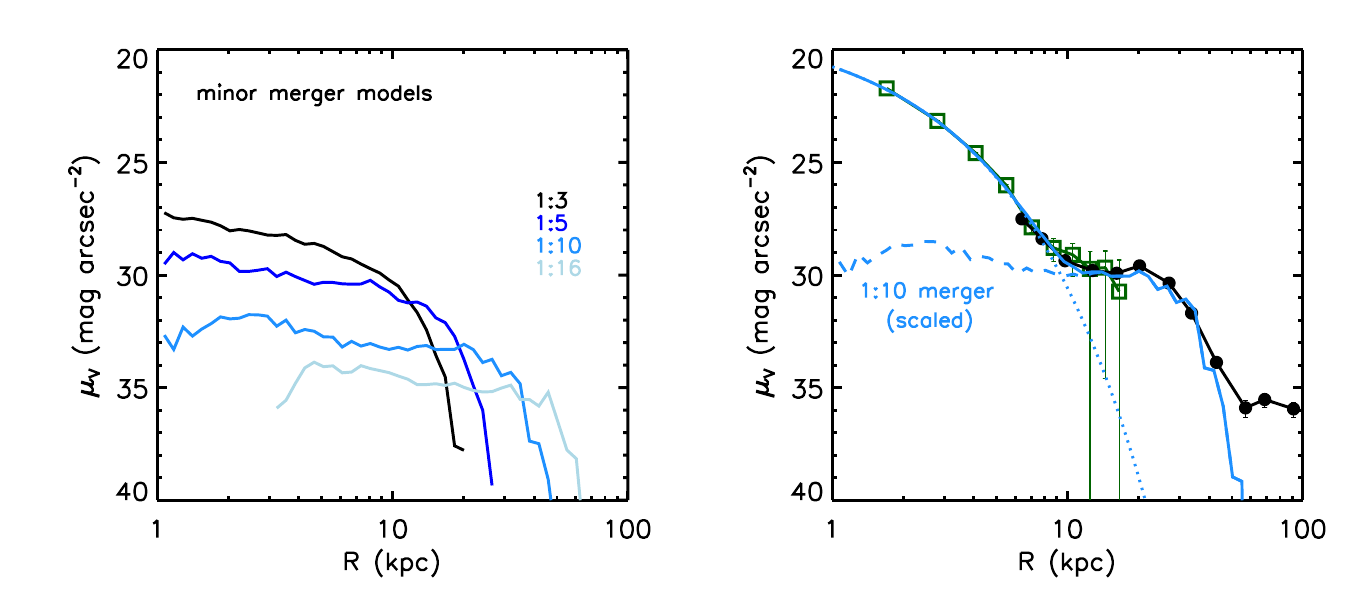}
\caption{{\it Left panel:} Surface brightness profiles of satellite
  debris resulting from minor mergers with merger ratios indicated in
  the figure.  Notice that higher mass ratios results in a more
  extended surface brightness profile.  {\it Right panel:} Comparison
  between the data and a model in which the host (represented by a
  dotted line) undergoes a 1:10 merger (dashed line); the combined
  model profile is shown as a solid blue line.  The merger is scaled
  up by a factor of 20 compared to the models in the left panel.
  Either Ark 227 underwent many such mergers, or the adopted stellar
  mass of the satellite merger was larger than assumed in the default
  model.}
\label{fig:models}
\end{figure*}

The left panel of Figure \ref{fig:models} shows the predicted surface
brightness profiles of the satellite debris for merger models in which
the total mass ratios are 1:3, 1:5, 1:10, and 1:16 (the total mass
includes baryonic and dark matter).  There are two important features
that vary with mass ratio: the slope of the surface brightness profile
becomes flatter for higher mass ratios, and the location of the sharp
cutoff extends to much larger radius for higher mass ratios.  The
origin of these trends is a consequence of the competing effects of
tidal stripping and dynamical friction \citep[see discussion
in][]{Amorisco17, Deason22}.  Dynamical friction scales with the mass
ratio, such that more equal mass mergers result in a much more
efficient sinking of the satellite to the center of the host, where
the satellite is then tidally stripped.  In contrast, for high mass
ratios, dynamical friction is inefficient, and so the satellite spends
much of its time in the outskirts of the host.  Tidal stripping still
occurs near the satellite orbit pericenter, but that material is then
able to travel to the outskirts, where it spends most of its time.
The properties of the initial satellite orbit have a surprisingly
small effect on the resulting surface brightness profile
\citep{Amorisco17}.  The largest effect is at small radius, which is
difficult to observe because of the overwhelming effect of the host
stellar density.

The right panel of Figure \ref{fig:models} shows a comparison between
the data and a merger model.  The model was constructed by combining a
by-eye Sersic fit to the inner light profile (dotted line) with a 1:10
total mass ratio merger model (dashed line).  The latter was scaled
upward in luminosity by a factor of 20.  This model does an excellent
job of reproducing the shelf in the surface brightness profile at
$10-50$ kpc.  The second, tentative, shelf at $>50$ kpc would seem to
require a merger with a total mass ratio $>$1:16, which is the most
extreme mass ratio we were able to simulate.  A lower satellite halo
concentration may also produce a more extended distribution
\citep{Amorisco17}.  Exploration of higher mass ratios and a wider
range of satellite properties is required to understand the nature of
the tenuous outer shelf in Ark227.

The observed shelf at $\sim10$ kpc is much more luminous than the
corresponding 1:10 total mass ratio merger.  We emphasize that the
model adds stars to the simulation by hand, and so the normalization
is not a strong prediction of the model, in contrast to the shape,
which {\it is} a strong prediction and is set by the total mass ratio
of the merger.  There are at least two possible explanations for the
large offset.  Ark 227 could have experienced many 1:10 mergers,
resulting in an aggregate luminosity comparable to the data.  This
seems unlikely because cosmological simulations do not predict such a
large number of 1:10 mergers.  It would also be difficult to imagine
such a large number mergers producing a very strong shelf feature.

A second possibility is that the satellite galaxy occupying the
$10^{10}\msun$ halo is more massive than assumed by Deason et al.  The
stellar mass in the shelf is $\approx 2\times10^8\msun$, or
approximately 10\% of the total stellar mass of Ark 227.  Deason et
al. adopted a fairly steep stellar mass-halo mass relation where
$M_\ast \propto M_h^{1.6}$.  Either this relation instead has
power-law index closer to $1.0$, or the satellite accreted by Ark 227
happens to be over-luminous for its halo mass.  With only a single
object it is difficult to reach a strong conclusion on this point.
Observations of additional dwarf halos are necessary to resolve this
issue.

\begin{figure*}[!t]
\center
\includegraphics[width=1.0\textwidth]{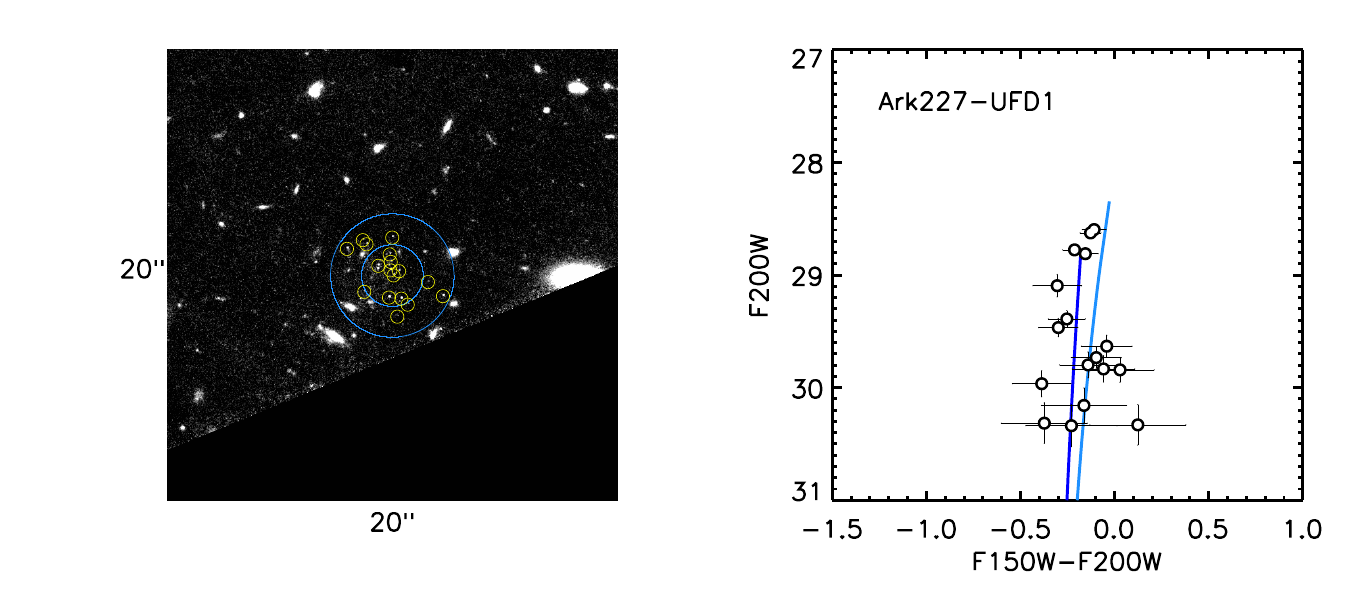}
\caption{The ultra-faint dwarf Ark227-UFD1.  {\it Left panel:}
  $20\arcsec$ cutout centered on the dwarf, which happens to fall
  toward the edge of the field.  Yellow circles show all sources
  satisfying the point-source selection in Figure \ref{fig:size-mag}.
  One and two times the half-light radius are indicated by the
  circles.  {\it Right panel:} CMD of all point sources within
  $2\arcsec$ of Ark227-UFD1.  The light and dark blue lines show
  isochrones for [Fe/H]$=-1.5$ and $-2.0$, respectively.  Ark227-UFD1
  has a half-light size of $1.4\arcsec\approx230$ pc, a surface
  brightness of 28.1 mag arcsec$^{-2}$ and a stellar mass of
  $\approx10^5\msun$.}
\label{fig:ufd}
\end{figure*}

\subsection{An ultra-faint dwarf galaxy associated with Ark 227}
\label{s:ufd}

Visual inspection of the spatial distribution of point sources
revealed a strong over-density of sources 50 kpc ($5\arcmin$) from the
center of Ark 227.  There are 17 sources within a few arcsec - a
spatial density far higher than the background stellar halo at this
projected separation.  

Figure \ref{fig:ufd} shows the spatial distribution (left panel) and
CMD (right panel) of these 17 sources.  Unlike the analysis in earlier
sections, a CMD filter has not been applied to this sample; the only
selection is a half-light size smaller than $0.01\arcsec$.  A strong,
roughly circular distribution of sources is clearly visible.  Notice
that there are no background halo stars within this
$20\arcsec \times 20\arcsec$ cutout.  The half-light radius of these
sources is $1.4\arcsec$ and is indicated by the smaller circle.  Twice
the half-light radius is indicated with a larger circle.  The right
panel shows the CMD along with isochrones for [Fe/H]$=-1.5$ and
$-2.0$.  The sources are clearly consistent with a metal-poor
population at the distance of Ark 227.  The surface brightness within
the half-light radius is $\mu_{V,e}=28.1$ mag arcsec$^{-2}$.

Assuming that these sources are associated with Ark 227, the physical
half-light size is 230 pc.  Summing up the flux from the 17 sources
and accounting for the unresolved flux from fainter sources implies a
luminosity of $M_V=-7.0$ and a stellar mass of
$M_\ast\approx10^5\msun$, assuming $M/L_V=2$.  The size and luminosity
are consistent with the properties of ultra-faint dwarf (UFD) galaxies
\citep{Simon19}; we therefore consider this object a UFD associated
with Ark 227 and refer to it as Ark227-UFD1.

We calculate the effective area covered by our two NIRCam pointings
and for which we could have detected a UFD.  We removed the lowermost
module from this estimation as there are too many point sources to be
able to easily identify an overdensity of stars associated with a
dwarf.  The source detection map was used to identify and mask large
galaxies from the effective area.  We find an effective area of
$\approx12$ arcmin$^2$.  The total area subtended by the halo virial
radius is 300 arcmin$^2$.  The effective area of our search represents
$\approx1/25$ of the total halo, suggesting that Ark 227 may harbor
several dozen UFDs at $M_\ast\sim10^5\msun$.

%--------------------------------------------------------%
%--------------------------------------------------------%
%--------------------------------------------------------%

\vspace{0.5cm}

\section{Summary \& Discussion}
\label{s:disc}

In this paper we reported the serendipitous discovery of at least one
-- and possibly two -- accretion shelves in the halo of the dwarf
galaxy Ark 227 ($M_\ast=5\times10^9\msun$;
$M_{\rm halo}\approx 2\times10^{11}\msun$).  Deep {\it JWST} NIRCam
imaging provided robust star-galaxy separation to
$m_{\rm AB}\approx30$, and enabled us to trace the surface brightness
profile of Ark 227 to a limit of $\mu_V\approx 35$ mag arcsec$^{-2}$
to this galaxy's predicted dark matter halo virial radius at 100 kpc.
One accretion shelf is clearly detected at $\mu_V\approx 30$ mag
arcsec$^{-2}$ at $10-20$ kpc from the center of Ark 227.  A second,
tentative shelf is detected at $\mu_V\approx 35$ mag arcsec$^{-2}$ at
$50-100$ kpc.  Stellar colors vs. radius provide evidence for abrupt
changes in metallicity at the location of these shelves.

Accretion shelves are generic predictions of hierarchical structure
formation \citep[e.g.,][]{Amorisco17, Deason22}.  Their amplitude and
location provide fairly direct information on the properties of the
accreted satellite, or satellites: the amplitude is determined by the
stellar mass ratio of the merger and the location by the halo mass
ratio.  In Ark 227, comparison to models suggests that Ark 227
experienced at least two minor mergers: a 1:10 merger with a galaxy of
stellar mass $10^8\msun$ and [Fe/H]$\approx-0.8$; and a >1:20 merger
with a galaxy of mass $10^7\msun$ and [Fe/H]$\approx-1.2$.  The
stellar masses and metallicities of these accreted dwarfs are
consistent with the observed mass-metallicity relation of intact
dwarfs measured in the local universe \citep{Kirby13}.  These deep
{\it JWST} data have enabled the most detailed reconstruction of the
hierarchical assembly of a dwarf galaxy to-date.

Stellar halos have been traced to the virial radii of the Milky Way
\citep{Deason18b}, M31 \citep{Ibata07}, and now Ark 227.  If we assume
that it is common for stars to populate the entire extent of dark
matter halos, we can estimate the fraction of the sky that is filled
with stellar halos.  For this estimate we use the empirical model from
\citet{Behroozi19}, which populates galaxies in a large cosmological
volume.  Halos hosting galaxies with log $M_\ast/\msun=9$ within 35
Mpc cover $5-15$\% of the sky.  Extending this to all halos above log
$M_\ast/\msun=8$ and within 70 Mpc, the covering fraction reaches
$20-30$\% of the sky.  These numbers imply that the existence of a
stellar halo in the foreground of one of the few well-studied
extragalactic deep fields, while surprising to us, is not an
exceptionally rare configuration.

With a stellar mass of $5\times10^{9}\msun$, Ark 227 lies at the upper
end of the dwarf galaxy mass scale.  \citet{Deason22} simulated
predicted stellar halos for galaxies with stellar masses of
$10^{7}\msun$, finding that the signatures of hierarchical assembly
may be present at the level of $\mu_V\sim35$ mag arcsec$^{-2}$.  This
limit has not yet been breached for very low mass dwarfs, but we have
shown here that {\it JWST} imaging is a unique and efficient tool for
such searches.  Future observations of nearby, isolated dwarf galaxies
with {\it JWST} should place strong constraints on the accretion
histories of low-mass galaxies.

%--------------------------------------------------------%

\acknowledgments 

This work is based on observations made with the NASA/ESA/CSA James
Webb Space Telescope. The data were obtained from the Mikulski Archive
for Space Telescopes at the Space Telescope Science Institute, which
is operated by the Association of Universities for Research in
Astronomy, Inc., under NASA contract NAS 5-03127 for JWST. These
observations are associated with program GO 1810.  The computations in
this paper were run on the FASRC Cannon cluster supported by the FAS
Division of Science Research Computing Group at Harvard University.

C.C. acknowledges support from JWST-GO-1810.  S.B. is supported by the
ERC Starting Grant “Red Cardinal”, GA 101076080.  R.P.N.  acknowledges
support for this work provided by NASA through the NASA Hubble
Fellowship grant HST-HF2-51515.001-A awarded by the Space Telescope
Science Institute, which is operated by the Association of
Universities for Research in Astronomy, Incorporated, under NASA
contract NAS5-26555.1.  R.E. acknowledges the support from grant
numbers 21-ATP21-0077, NSF AST-1816420, and HST-GO-16173.001-A as well
as the Institute for Theory and Computation at the Center for
Astrophysics.

%--------------------------------------------------------%

%\bibliography{../master_refs}

\end{document}